\documentclass{moriond}
\usepackage{amssymb}
\usepackage{amsthm,enumerate}
\usepackage[all]{xy}

\def\Journal#1#2#3#4{{#1} {\bf #2}, #3 (#4)}

\def\PRL{\em Phys. Rev. Lett.}
\def\PRD{{\em Phys. Rev.} D}

\newcommand{\Photo}{\includegraphics[height=35mm]{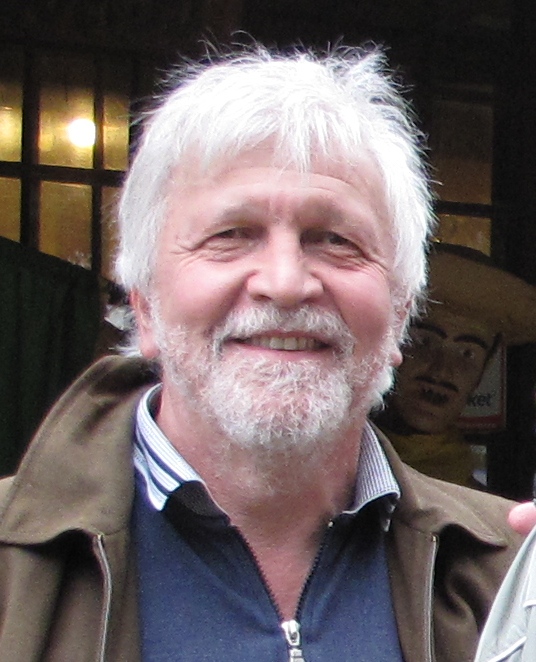}}

\begin{document}
\vspace*{4cm}
\title{RANDALL-SUNDRUM  VS. HOLOGRAPHIC BRANEWORLD}

\author{NEVEN BILIC}

\address{Division of Theoretical Physics, Rudjer Bo\v skovi\'c Institute, Bijeni\v cka 54,\\
Zagreb, Croatia}

\maketitle
\abstracts{
A mapping between two braneworld cosmologies -- Randall-Sundrum and holographic 
 -- is explicitly constructed. The cosmologies are governed by the appropriate modified Friedman equations.
 A relationship between the corresponding Hubble rates is established.
}

\section{Introduction}

Braneworld cosmology is based on the scenario in which matter is confined on a brane moving in 
the higher dimensional bulk with only 
gravity allowed to propagate in the bulk.
We will consider two types of braneworlds in an AdS$_5$ bulk.
In a holographic braneworld universe a 3-brane is located 
at the boundary of the asymptotic AdS$_5$ . The cosmology is
governed by matter on the brane in addition to the
boundary CFT. In the second Randall-Sundrum (RSII) model \cite{randall} a 3-brane is
located at a finite distance from the boundary of AdS$_5$.
The model was originally proposed as an alternative to
compactification of extra dimensions.

A cosmology on the brane is obtained by allowing the brane to
move in the bulk along the fifth dimension $z$. Equivalently, the brane is kept fixed at $z = z_{\rm br}$
while making the metric in the bulk time dependent.
The time dependent bulk spacetime with line element  
\begin{equation}
ds_{(5)}^2=\frac{\ell^2}{z^2}\left( 
n^2(\tau,z)d\tau^2- a^2(\tau,z) d\Omega_\kappa^2-dz^2 
\right) ,
 \label{eq102}
\end{equation}
may be regarded as
a $z$ foliation of the bulk with an FRW cosmology on each $z$ slice.
In particular, 
at $z=z_{\rm br}$ we have the RSII cosmology and, at z=0, the holographic cosmology.
The Friedmann equation on the brane is modified
\begin{equation}
H^2+\frac{\kappa}{a^2}=\frac{8\pi G_{\rm N}}{3} \rho +
\left(\frac{4\pi G_{\rm N}\ell}{3}\right)\rho^2
+\frac{\mu\ell^2}{a^4} ,
\label{eq022}
\end{equation}
where $\ell$ is the AdS$_5$ curvature radius, $H=\dot{a}/(na)$ is the Hubble rate
and $\mu$ is the parameter related to the bulk black-hole mass
$\mu=(8G_5 M_{\rm bh})/(3\pi \ell^2$.

In the RSII model by introducing the boundary in AdS$_5$ at
$z = z_{\rm br}$ instead of $z = 0$, the model is conjectured to be dual to
a cutoff CFT coupled to gravity,
with $z=z_{\rm br}$ providing
the IR cutoff. 
This conjecture  then reduces to
the standard AdS/CFT duality as the boundary is pushed off to $z=0$. This
connection involves a single CFT at the boundary  of a single patch of AdS$_5$.
In the original  RSII model one assumes
the $Z_2$ symmetry  $z\leftrightarrow z_{\rm br}^2/z$,  so
the region $0<z\leq z_{\rm br}$ is identified with $z_{\rm br} \leq z <\infty$, 
with the observer brane at the fixed point $z=z_{\rm br}$.
Hence, the braneworld is sitting between two patches of AdS$_5$, one on either side, and is
therefore dubbed ``two sided''.
In contrast, in the ``one-sided'' 
 RSII model the region $0\leq z\leq z_{\rm br}$ is simply cut off 
so the bulk is the section of spacetime $z_{\rm br} \leq z <\infty$.


The variation of the action yields Einstein's equations on the
boundary \cite{bilic}
\begin{equation}
R_{\mu\nu}- \frac12 R g_{\mu\nu}= 8\pi G_{\rm N} (\gamma \langle T^{\rm CFT}_{\mu\nu}\rangle +T^{\rm matt}_{\mu\nu}),
 \label{eq3006}
\end{equation}
where 
 $T^{\rm matt}_{\mu\nu}$ is the energy-momentum tensor associated with matter on the holographic brane 
and  $T^{\rm CFT}_{\mu\nu}$ the energy-momentum tensor of
the CFT on the boundary.
The parameter $\gamma$ takes on the value 1 and 2 for the 1-sided and 2-sided  RSII model, respectively.
According to the AdS/CFT prescription, the expectation value $\langle T^{\rm CFT}_{\mu\nu}\rangle$
is obtained 
by functionally differentiating the renormalized on-shell bulk gravitational action with respect to the
boundary metric \cite{haro}. 
From (\ref{eq3006}) one derives the Friedmann equation at the boundary
\begin{equation}
 \mathcal{H}_0^2=\frac{\ell^2}{4}
 \left(\mathcal{H}_0^4+ \frac{4\mu}{a_0^4} \right)+\frac{8\pi G_{\rm N}}{3}\rho_0.
 \label{eq3110}
\end{equation}
where $\mathcal{H}_0^2=H_0^2+\kappa/a_0^2$ and $H_0=\dot{a}_0/a_0$ is the Hubble rate
at the holographic boundary.

A map between $z$-cosmology and $z=0$-cosmology can
be constructed using \cite{apostolopoulos}
\begin{equation}
a^2=a_0^2\left[
\left(1-\frac{\mathcal{H}_0^2 z^2}{4}\right)^2
+ \frac14 \frac{\mu z^4}{a_0^4}
\right],
\quad
n=\frac{\dot{a}}{\dot{a}_0}.
 \label{eq3103}
\end{equation}
The Hubble rates are related by
\begin{equation}
\mathcal{H}\equiv H^2+\frac{\kappa}{a^2} = \mathcal{H}_0\frac{a_0}{a}.
 \label{eq201}
\end{equation}
Using this and (\ref{eq3103}) we can find a relation between the cosmological scales $a_{\rm br}$  on the
brane at $z=z_{\rm br}$ and $a$ at on an arbitrary $z$-slice.
First, we can express the first equation in (\ref{eq3103}) as an equation for $a_0^2$, $a^2$, and $\mathcal{H}^2$, 
and similarly as another equation for $a_0^2$, $a_{\rm br}$, and $\mathcal{H}_{\rm br}^2$   $z=z_{\rm br}$.
By eliminating $a_0^2$ from  these
two equations  we find 
\begin{equation}
a=\frac{a_{\rm br}}{\sqrt2}
\left[\left(1 +\frac12 \mathcal{H}_{\rm br}^2 z_{\rm br}^2\right)\left(1+\frac{z^4}{z_{\rm br}^4}  \right)
- \mathcal{H}_{\rm br}^2 z^2 
+\mathcal{E} (z) 
\sqrt{1+\mathcal{H}_{\rm br}^2 z_{\rm br}^2
-\frac{\mu z_{\rm br}^4}{a_{\rm br}^4}}\left(1-\frac{z^4}{z_{\rm br}^4}  \right)
\right]^{1/2} ,
 \label{eq3205}
\end{equation}
where we have introduced a two-valued step function
\begin{equation}
\mathcal{E}(z)=\left\{ \begin{array}{ll}
+1, & \mbox{ for $z\geq z_{\rm br}$},\\
-1,& \mbox{ for $z<z_{\rm br}$, two-sided version},\\
+1 \mbox{ or } -1, & \mbox{ for $z<z_{\rm br}$, one-sided version} . \end{array} \right.
\label{eq4105}
\end{equation}
The map is schematically illustrated as
$$
\xymatrix{
d\tau^2 -a_0^2 d\Omega_\kappa^2  \ar[rr]^{\tau\rightarrow\tilde{\tau}} \ar[d]_z & & 
(1/n^2)d\tilde{\tau}^2 -a_0^2 d\Omega_\kappa^2 \ar[d]^z \\
n^2 d\tau^2 -a^2 d\Omega_\kappa^2 \ar[rr]_{\tau\rightarrow\tilde{\tau}} & & 
d\tilde{\tau}^2 -a^2 d\Omega_\kappa^2
}
$$
where  $\tau$ and $\tilde{\tau}$ are the holographic and RSII synchronous times, respectively.
By making use of  (\ref{eq201}) and (\ref{eq3205}) we express the Hubble rate at $z=0$ in terms of
the Hubble rate at $z=z_{\rm br}$
\begin{equation}
\mathcal{H}_0^2= 2\mathcal{H}_{\rm br}^2
\left(
1 +\frac{\mathcal{H}_{\rm br}^2 z_{\rm br}^2}{2} +\mathcal{E}_0
\sqrt{1+\mathcal{H}_{\rm br}^2 z_{\rm br}^2
-\frac{\mu z_{\rm br}^4}{\mathcal{A}_{\rm br}^4}}
\right)^{-1} ,
 \label{eq3209}
\end{equation}
where $\mathcal{E}_0\equiv \mathcal{E}(0)=-1$ for the two-sided and $\mathcal{E}_0=+1$ or $-1$ for the one-sided version
of the RSII model.
There is a clear distinction between the holographic maps involving 1-sided and 2-sided versions of
the RSII model \cite{bilic}.
In the 2-sided map the low-density regime on 
the RSII brane corresponds to the high negative energy density on
the holographic brane.
The low density regime can be made simultaneous only in the
1-sided RSII.

It is conceivable that we live in a braneworld with emergent
cosmology. That is, dark energy and dark matter could be
emergent phenomena induced by what happens on the primary
braneworld.
For example, suppose our universe is a one-sided RSII
braneworld the cosmology of which is emergent in parallel
with the primary holographic cosmology. If $\rho_0$ describes matter
with the equation of state satisfying $3p_0 +\rho_0 >0$, as for, e.g.,
CDM, we will have an asymptotically de Sitter universe on the
RSII brane. If we choose $\ell$ so that the cosmological constant $\Lambda$ fits the observed value, the
quadratic term will be comparable with the linear term today but
will strongly dominate in the past and hence will spoil the
standard cosmology. However, the standard $\Lambda$CDM cosmology
could be recovered by including a negative 
constant term in $\rho_0$ and fine tune it to cancel $\Lambda$ up to a small
phenomenologically acceptable contribution.
\section*{Acknowledgments}
This work has been supported by the Croatian 
Science Foundation, Project No.\ IP-2014-09-9582.

\end{document}